\documentclass[aps,prd,twocolumn,floatfix,nofootinbib,showpacs,superscriptaddress,tightenlines]{revtex4}

\usepackage{amsmath}
\usepackage{dcolumn}
\usepackage{lipsum}
\usepackage{amssymb}
\usepackage[utf8]{inputenc}
\usepackage{url}
\usepackage{epsfig}
\usepackage{graphicx}
\usepackage{amsmath}
\usepackage{bm}
\usepackage{setspace}
\usepackage{lscape}
\usepackage{amsthm}
\usepackage{bbold}
\usepackage{dcolumn}
\usepackage{epsfig}
\usepackage{graphics}
\usepackage{graphicx}
\usepackage{longtable}
\usepackage{color}
\usepackage{xcolor}
\usepackage{bm}
\usepackage{xspace}
\usepackage{cancel}
\usepackage{float}
\usepackage{multirow}
\usepackage{soul}
\usepackage[mathscr,scaled=1.15]{urwchancal}
\DeclareFontFamily{OT1}{pzc}{}
\DeclareFontShape{OT1}{pzc}{m}{it}%
{<-> s * [1.15] pzcmi7t}{}
\DeclareMathAlphabet{\mathpzc}{OT1}{pzc}{m}{it}

\definecolor{darkgreen}{rgb}{0,0.5,0}
\definecolor{amber}{rgb}{1.0, 0.75, 0.0}
\definecolor{purple}{rgb}{0.5,0,0.5}
\definecolor{nblue}{rgb}{0.0,0.0,0.50}
\definecolor{scarlet}{rgb}{1.0,0.2,0}
\definecolor{darkmagenta}{rgb}{0.55, 0.0, 0.55}
\definecolor{darkolivegreen}{rgb}{0.33, 0.42, 0.18}
\definecolor{darkcandyapplered}{rgb}{0.64, 0.0, 0.0}
\definecolor{warmblack}{rgb}{0.0, 0.26, 0.26}
\definecolor{oxfordblue}{rgb}{0.0, 0.13, 0.28}
\definecolor{cyan(process)}{rgb}{0.0, 0.55, 0.55}
\definecolor{almond}{rgb}{0.94, 0.87, 0.8}
\definecolor{antiquewhite}{rgb}{0.98, 0.92, 0.84}
\definecolor{eggshell}{rgb}{0.94, 0.92, 0.84}
\definecolor{floralwhite}{rgb}{1.0, 0.98, 0.94}
\definecolor{linen}{rgb}{0.98, 0.94, 0.9}
\definecolor{darkred}{rgb}{0.55, 0.0, 0.0}
\usepackage[colorlinks=true, pdfstartview=FitV, linkcolor=purple, citecolor= purple, urlcolor=blue]{hyperref}

\newcommand{\be}{\begin{equation}}
\newcommand{\tu}{\textcolor{red}{u}}

\newcommand{\td}{\textcolor{darkcandyapplered}{d}}
\newcommand{\tb}{{\textcolor{blue}{b}}}
\newcommand{\tc}{{\textcolor{darkred}{c}}}
\newcommand{\ts}{{\textcolor{darkgreen}{s}}}
\newcommand{\ee}{\end{equation}}
\newcommand{\bea}{\begin{eqnarray}}
\newcommand{\eea}{\end{eqnarray}}
\newcommand{\beas}{\begin{eqnarray*}}
\newcommand{\eeas}{\end{eqnarray*}}
\newcommand{\nn}{\nonumber}

\newcommand{\GeV}{\text{GeV}} %math GeV

\begin{document}
\title{Mass Spectra of One or Two Heavy Quark Mesons  and Diquarks within a non-relativistic potential model}

\author{L.X. Guti\'errez-Guerrero}
\email[]{lxgutierrez@mctp.mx}
\thanks{}
\affiliation{CONACyT-Mesoamerican Centre for Theoretical Physics,
Universidad Aut\'onoma de Chiapas, Carretera Zapata Km. 4, Real
del Bosque (Ter\'an), Tuxtla Guti\'errez 29040, Chiapas, M\'exico.}
       
\author{Jes\'us Alfaro}
\email[]{jesussk\_28@hotmail.com }
\thanks{}
\affiliation{Facultad de Ciencias en F\'isica y Matem\'aticas, Universidad Aut\'onoma de Chiapas (FCFM-UNACH),
Ciudad Universitaria UNACH, Carretera Emiliano Zapata Km. 8 Rancho San Francisco, Ciudad
Universitaria Ter\'an, Tuxtla Guti\'errez, Chiapas C.P. 29050 Chiapas, M\'exico,}

\author{A. Raya}
\email[]{alfredo.raya@umich.mx}
\thanks{}
\affiliation{Instituto de F\'isica y Matem\'aticas, Universidad
Michoacana de San Nicol\'as de Hidalgo, Edificio C-3, Ciudad
Universitaria, Morelia, Michoac\'an 58040, M\'exico.}
\affiliation{Centro de Ciencias Exactas, Universidad del Bío-Bío,\\ Avda. Andrés Bello 720, Casilla 447, 3800708, Chillán, Chile.}

\begin{abstract}
In this article, the mass spectra of mesons with one or two heavy quarks and their diquarks partners are estimated within a non-relativistic framework by solving Schr\"odinger equation with an effective potential inspired by a symmetry preserving Poincaré covariant vector-vector contact interaction model of quantum chromodynamics. Matrix Numerov method is implemented for this purpose. In our survey of mesons with heavy quarks, we fix the model parameter to the masses of ground-states and then extend our calculations for radial excitations and diquarks. The  potential model used in this work gives results which are in good agreement with experimental data and other theoretical calculations.

\end{abstract}

\maketitle

%\section{Introduction}
\section{Introduction}

A major challenge in hadron physics is a complete understanding of the dynamics of mesons and baryons with constituent charm and bottom quarks.
The discovery of the $J/\psi$ meson in 1974~\cite{Aubert:1974js,Augustin:1974xw}, opended the door for the observation of other charmonium states ~\cite{Appelquist:1974yr,Swanson:2006st,Lebed:2016hpi}.
%%%%%%%%%%%%%%%%%%%%%%%%%%%%%%%%%%%%%%%%
At the same time, spectroscopy of hadrons with $\tb$ quarks has represented a fundamental tool to understand QCD, and for this reason, several experiments in Fermilab, CERN, and other facilities continue to orient some of their scientific work to this problem.
%%%%%%%%%%%%%%%%%%%%%%%%%%%%%%%%%%%%%%%%%
The family of $B_c$ mesons with $\tb$ and $\tc$ quarks have a privileged place in hadron physics. Its mass is found in a range between the corresponding to charmonium and bottomonium states. $B_c$ meson is the only state composed of two different flavors of valence quarks. Therefore, this state provides a special way to explore the heavy quark dynamics that are complementary to those provided by the $\tc\bar{\tc}$ and $\tb\bar{\tb}$ states. The ground-state of $B_c$ meson was first observed in 1998 at Fermilab in Collider Detector~\cite{Abe:1998wi} with a mass of 6.2749~GeV. The $B_\tc$ mass spectrum gives valuable information about heavy-quark dynamics and improves the understanding of the behavior of fundamental interactions. It has been obtained within different models~\cite{Eichten:1994gt,Eichten:2019gig,Li:2019tbn,Chen:2020ecu,Akbar:2019kbi}, including some lattice simulations~\cite{Mathur:2018epb,Dowdall:2012ab,Davies:1996gi}. From the experimental perspective,  observation of $B_\tc$ mesons demands the production of both $\tc \bar{\tc}$ and $\tb\bar{\tb}$ pairs.Thus,  the production rate is small and as a consequence, these mesons have been less studied than charmonium or bottomonium.\\
%%%%%%%%%%%%%%%%%%%%%%%%%%%%%%%%%%%%%%%%%%%%%%%%%%%%%%%
In hadron physics, mesons composed of one heavy and one light quarks might be regarded as analogs of the hydrogen atom in the sense that relativistic effects are reduced and the heavy quark acts as a pointlike probe (nucleus) of the light constituent quarks (electron)~\cite{Flynn:1992fm,Godfrey:2016nwn}.  
In this sense, a non-relativistic approach to explore the mass spectra of these object might be justified because the heavy quark, either charm or bottom, have masses of about 1800~MeV and 5300~MeV and thus are non-relativistic in the sense that binding effects are small compared to these masses. For light quarks, this argument should be taken with a grain of salt. One might observe the behavior of the velocity of the quark inside the hadron compared with the momentum width of the state. If such a velocity scales with the inverse of the quark mass or the bound state mass we have means to distinguish whether the state  under consideration might be regarded as relativistic or non-relativistic.~\cite{alfredo2021}.

Heavy-Light hadrons also provide an ideal platform to study non-perturbative phenomena of QCD~\cite{Lu:2016bbk,DeRujula:1976ugc,Rosner:1985dx}, as they are the bridge between the lightest and heaviest hadrons. In recent years, several Heavy-Light mesons have been observed in experiments~\cite{Zyla:2020zbs}, and thus, have attracted the attention of scientists due to their difference with light mesons. These observed meson states could be related to diquarks. Let us recall that the
concept of diquarks is not new. Gell-Mann~\cite{GellMann:1964nj} predicted the existence of these states in his quark model. Soon after, the idea was adapted to describe baryons as composed of a constituent diquark and a quark~\cite{Ida:1966ev,Lichtenberg:1967zz}. 
This simplification allows addressing the three-body problem as a two-body system. Non-point-like diquarks are crucial in hadron physics~\cite{Barabanov:2020jvn} and have been incorporated in new theoretical techniques for the continuum bound-state problem and lattice-regularised QCD. Also, diquark are essential for studying the bound states of multiquark states.
Diquark masses are often calculated via phenomenological considerations~\cite{Wilczek:2004im,Selem:2006nd}; or can be predicted by binding two quarks via a one-gluon-exchange interaction term~\cite{Anwar:2017toa,Maiani:2015vwa} plus spin-spin corrections~\cite{Maiani:2004vq}.\\
%%%%%%%%%%%%%%%%%%%%%%%%%%%%%%%%%%%%%%%%
All heavy quark systems can be described in non-relativistic terms by modeling the strong interactions through an effective potential of the inter-quark separation which then is introduced in a Schrödinger equation to describe the mass spectra of this states~\footnote {A word of caution has to be addressed for light quark bound states, as the non-relativistic description might not be fully justified.}.  A prototypical example is the Cornell potential and its extensions, for which meson-bound states have widely been investigated. 
%%%%%%%%%%%%%%%%%%%%%%%%%%%%%%%%%%%%%%%%%%%%%%%%%%%%%%%
Solutions of the Schrödinger equation with spherically symmetric potentials play an important role in many fields of physics and hadronic spectroscopy is not an exception. A variety of numerical techniques are readily available in the literature, including Shooting~\cite{nrecipes} and the Asymptotic Iteration~\cite{Mutuk:2018dtg} methods. \\
%%%%%%%%%%%%%%%%%%%%%%%%%%%%%%%%%%%%%%%%%%%%%%%%%%%%%%%%
In this work, we present the calculation of heavy meson mass spectra using an effective semi-relativistic potential inspired on a symmetry preserving Poincaré covariant vector-vector contact-interaction (CI) model of QCD. This formalism was first proposed in~\cite{GutierrezGuerrero:2010md}. Later, it was used to study many light hadrons~\cite{GutierrezGuerrero:2010md,Roberts:2010rn,Roberts:2011wy,Chen:2012qr,Xu:2015kta}.  Charmed and bottom mesons using this approach were predicted in ~\cite{Bedolla:2015mpa,Serna:2017nlr,Raya:2017ggu,Yin:2019bxe}. the spectrum of the strange and nonstrange hadrons with the CI treatment was published in~\cite{Chen:2012qr}. Parity partners with this model were presented in~\cite{Lu:2017cln}.  A unified picture from masses of light and heavy mesons and baryons is given in~\cite{Yin:2019bxe,Chen:2019fzn,Gutierrez-Guerrero:2019uwa}. The purpose of this article is to provide an alternative description of the mass spectra of Heavy-Heavy and Heavy-Light mesons and their associated diquarks from a non-relativistic approach. We have organized the remaining of the work as follows. In Section~\ref{Numerov} we discuss the Matrix Numerov Method (MNM) to solve the one-body reduced Schrödinger equation. Section~\ref{Potential} contains details on the potential used in this work to calculate the mass spectra of mesons and diquarks. We obtain the masses of bound states with one heavy quark and one light quark and its radial excitations in  Section~\ref{HL-Mesons}. Then, we calculate the masses of the heaviest mesons and diquarks, together with their radial excitations. Section~\ref{conclusions} presents a summary and also a perspective on extensions of this study, including new possible directions.

%%%%%%%%%%%%%%%%%%%%%%%%%%%%%%%%%%%%%%%%%%%%%%%%%%%%%%%
%%%%%%%%%%%%%%%%%%%%%%%%%%%%%%%%%%%%%%%%%%%%%%%%%%%%%%%
%%%%%%%%%%%%%%%%%%%%%%%%%%%%%%%%%%%%%%%%%%%%%%%%%%%%%%%
\section{Matrix Numerov Method}
\label{Numerov}
%%%%%%%%%%%%%%%%%%%%%%%%%%%%%%%%%%%%%%%%%%%%%%%%%%%%%%%%%%%%%%%%%%%%%%%%%%%%%%%%%%%%%%%%%%%%%%%%%%%%%%%%%%%%%%%%%%%%%%%%%%%%%%%%%%%%%%%%%%%%%%%%%%%%%%%%%%%%%%%%%%%%%
In non-relativistic quantum mechanics, the basic object of study is the Schrödinger equation. Its stationary form corresponds to the eigenvalue equation
\begin{equation}\label{1}
\mathscr{H}\psi(r,\theta,\phi)=\Delta_M\psi(r,\theta,\phi),\end{equation}
where $\psi$ is the wave function, $\mathscr{H}=\mathscr{T}+\mathscr{V}$ is the Hamiltonian of the system with kinetic energy $\mathscr{T}$, potential $\mathscr{V}$ and $\Delta_M$ is the eigenvalue which in this case corresponds to the mass of the bound state. %In the formalism used herein, $\Delta$  represents the mass of the bound state. 
In the non-relativistic regime, the kinetic energy term is expressed as
\bea\mathscr{T}=m_{Q}+m_{\overline{q}}+\frac{1}{2\mu}(p_x^2+p_y^2+p_z^2),\eea
$m_Q$ and $m_{\overline{q}}$ representing the quark and antiquark masses, respectively, and $\mu=m_Qm_{\overline{q}}/(m_Q+m_{\overline{q}})$ is the reduced mass of the system. Since the potential depends only on the relative distance, we use the center of mass $(\mathbf{R})$ and relative distance $(\mathbf{r})$ vectors,
\bea
\mathbf{R}=\frac{ m_1\mathbf{r}_1 + m_2\mathbf{r}_2}{m_1+m_2},\;\;\;\mathbf{r}= \mathbf{r}_1- \mathbf{r}_2 ,\eea
and thus the wave function can be written as the product $\psi(\mathbf{R},\mathbf{r})=\varphi (\mathbf{R})\rho (\mathbf{r})$. Neglecting the center of mass contribution, we focus in the relative motion alone. Considering a spherically symmetric potential, the radial solution can be set as $U(r) = r\rho(r)$, such that the one-body radial equation becomes
\bea \nn\label{SE}
&&\left[-\frac{1}{2\mu}\frac{d^2}{dr^2}+\frac{l(l+1)}{2\mu r^2}+V(r)+m_Q+m_{\overline{q}}\right] U(r)\\
&&=\Delta_M U(r),\eea
where the second term represent the centrifugal barrier and $l$ is the orbital angular momentum of the meson.  MNM is a specialized strategy for numerically integrating differential equations of the form~\cite{doi:10.1119/1.4748813}
\bea
U''(r)=f(r)U(r)
\eea
through the rule
$$U_{i+1}=\frac{U_{i-1}(12-d^2f_{i-1})-2U_i(5d^2f_i+12)}{d^2f_{i+1}-12},$$
where $x_i$ denote a set of equidistant point separated a distance $d$, $U_{i}=U(x_i)$ and $f_i=f(x_i)$ such that the original differential equation can be written in matrix form as 
\be
A_{N,N}\vec{U}=B_{N,N}\vec{f}\vec{U},
\ee
where 
\bea A_{N,N}=\frac{I_{-1}-2I_0+I_1}{d^2},\quad
B_{N,N}=\frac{I_{-1}+10I_0+I_1}{12},\eea
where $I_{-1}$, $I_0$ and $I_1$ are,
respectively, the sub-, main- and up-diagonal unit matrices and the components of the vectors $\vec U$ and $\vec{f}$ correspond to the evaluation of the respective functions at each of the grid points.

In our case, we have
\bea f(r)=-2\mu\left[\Delta_M-V(r)-\frac{l(l+1)}{2\mu r^2}-m_Q-m_{\overline{q}}\right].\eea
%
%The method consists in defining an interval of integration $r \in (0, r_{max})$ and construct an equidistant set of $N$ points $r_i$ separated a distance $d$ 
such that Eq. (\ref{SE}) can be cast in the matrix form
\bea
-\frac{1}{2\mu}A_{N,N}B_{N,N}^{-1}\psi_i+\left[\frac{l(l+1)}{2\mu r_{i}^2}+V_N\right]\psi_i=\Delta_M\psi\;, \eea
where $\psi_i=\psi(r_i)$ %and the matrices
%%%%%%%%%%%%%%%%%%%%%%%%%%%%%%%%%%%%%%%%%%%%%%%%%%%%%%%%%%%%%%%%%%%%
%
and $V_N(r)=diag(\dots,V_{i-1},V_i,V_{i+1},\dots)$, with $V_i=V(r_i)+m_Q+m_{\overline{q}}$.\\
%%%%%%%%%%%%%%%%%%%%%%%%%%%%%%%%%%%%%%%%%%%%%%%%%%%%%%%%%%%%%%%%%%%%
We employ this formalism to calculate the masses Heavy-Heavy and Heavy-Light mesons  and their associated diquarks. Its applicability is restricted by the analytic form  of the potential, which we describe below. 
%%%%%%%%%%%%%%%%%%%%%%%%%%%%%%%%%%%%%%%%%%%%%%%%%%%%%%%%%%%%%%%%%%%%%%%%%%%%%%%%%%%%%%%%%%%%%%%%%%%%%%%%%%%%%%%%%%%%%%%%%%%%%%%%%%%%%%%%%%%%%%%%%%%%%%%%%%%%%%%%%%%%%%%%%%%%%%%%%%%%%%%%%%%%%%%%%%%%%%%%%%%%%%%%%%%%%%%%%%%%%%%%%%%%%%%%%%%%%%%%%%%%%%%%%%%%%%%%%%%%%%%%%%%%%%%%%%%%%%%%%%%%%%%%%%%%%%%%%%%%%%%%%%%%%%%%%%%%%%%%%%%%%%%%%%%%%%%%%%%%
\section{Effective Potential Model}
\label{Potential}
%%%%%%%%%%%%%%%%%%%%%%%%%%%%%%%%%%%%%%%%%%%%%%%%%%%%%%%%%%%%%%%%%%%%%%%%%%%%%%%%%%%%%%%%%%%%%%%%%%%%%%%%%%%%%%%%%%%%%%%%%%%%%%%%%%%%%%%%%%%%%%%%%%%%%%%%%%%%%%%%%%%%%%%%%%%%%%%%%%%%%%%%%%%%%%%%%%%%%%%%%%%%%%%%%%%%%%%%%%%%%%%%%%%%%%%%%%%%%%%%%%%%%%%%%%%%%%%%%%%%%%%%%%%%%%%%%%%%%%%%%%%%%%%%%%%%%%%%%%%%%%%%%%%%%%%%%%%%%%%%%%%%%%%%%%%%%%%%%%%%
Inspired by the vector-vector CI model proposed in~\cite{GutierrezGuerrero:2010md}, we start from the following form for the dressed-gluon propagator~\cite{Chang:2006bm}
\bea\label{gp}\Delta_{\mu\nu}(k)=\displaystyle\frac{1}{m_G^2}\delta_{\mu\nu}\Theta(\Lambda^2-k^2)\Theta(k^2-\lambda^2),\eea
where $\lambda$ and $\Lambda$ serve as cut-offs, and
\bea
m_G=\sqrt{\frac{m_g^2}{4\pi\alpha}},\eea
with $m_g$ is a gluon mass scale generated dynamically in QCD~\cite{Boucaud:2011ug,Aguilar:2017dco,Binosi:2017rwj,Gao:2017uox}. The parameter $\alpha$ can be interpreted as the interaction strength in the infrared~\cite{Binosi:2016nme,Deur:2016tte,Rodriguez-Quintero:2018wma}. Studies based on CI show that gluon propagator in eq.~({\ref{gp}}) provides an accurate description of masses and other static quantities for hadrons~\cite{GutierrezGuerrero:2010md,Roberts:2010rn,Roberts:2011cf} which can be used as bounds of full QCD calculations.
In order to describe bound states of heavy quarks within the non-relativistic framework,  we need an effective form of the potential. In the static limit, $V(\mathbf{r})$ can be obtained as the Fourier transform of the gluon propagator~\cite{Cucchieri:2017icl}, namely,
\bea
V_0(\mathbf{r})=\int \frac{d^3k}{(2\pi)^3}\;e^{ir\cdot k}\;\Delta_{00}(\mathbf{k}).\label{eqn1}\eea
In our case, taking the magnitudes $r=|\mathbf{r}|$ and $k=|\mathbf{k}|$, we straightforwardly obtain
\bea
\label{potential-CI}\nn
V_0(r)=\frac{1}{2\pi^2 m_G^2r^3}\left(\lambda
r\cos(\lambda r)-\Lambda r\cos(\Lambda r)\right.\\
\left. +\sin(\Lambda
r)-\sin(\lambda r)\right).\eea
The potential thus obtained is used to calculated hadrons containing one or two heavy quarks by replacing  $V(r)$ for the above $V_0(r)$ in Eq.~(\ref{SE}). Relativistic spin-spin and spin-orbit corrections are introduced by recalling that there are five long-living quark flavours that produce measurable bound states, namely, $\tu,\td,\ts,\tc,\tb$, out of these bound states which also include baryons,  mesons and diquarks of our interest  can have spin zero or spin one. %Table~\ref{table-mesones-lorentz} summarizes the Lorentz structure of quark interactions in the nonrelativistic limit. 
We specialize in the calculation of  pseudoscalar and vector mesons along with  scalar and axial vector diquarks, for which the spin-orbit term never appears.
%%%%%%%%%%%%%%%%%%%%%%%%%%%%%%%%%%%%%%%%%%%%%%%%%%%%%%%%%%%%%%%%%%%%%%%%%%%%%%%%%%%%%%%%%%%%%%%%%%%%%%%%%%%%%%%%%%%%%%%%%%%%%%%%%%%%%%%%%%%%%%%%%%%%%%%%%%%%%%%%%%%%%
%\vspace{-1.8cm}
%\begin{table}[H]
%\caption{\label{table-mesones-lorentz} Nonrelativistic interaction potential from the various conceivable Lorentz structures of an arbitrary fermion-antifermion interaction. }
%\begin{tabular}{@{\extracolsep{0.5 cm}}ccc}
%\hline
%\hline
%Name & Lorentz structure & Potential\\
% \rule{0ex}{3.5ex}
%Scalar & $1\otimes 1$ &$V_{S}=V_0+\langle H_{SS}\rangle$\\
% \rule{0ex}{3.5ex}
%Peudoscalar & $\gamma_5\otimes \gamma^5$& $V_0+\langle H_{SS}\rangle $\\
%\rule{0ex}{3.5ex}
%Vector &  $\gamma_\mu\otimes \gamma^\mu$ &$V_0+\langle H_{SS}\rangle$\\
%\rule{0ex}{3.5ex}
%Axial-Vector &  $\gamma_\mu\gamma_5\otimes \gamma^\mu\gamma^5 $& $V_0+\langle H_{SS}\rangle+\langle H_{LS}\rangle$\\
%\hline
%\hline
%\end{tabular}
%\end{table}
%%%%%%%%%%%%%%%%%%%%%%%%%%%%%%%%%%%%%%%%%%%%%%%%%%%%%%%%%%%%%%%%%%%%%%%%%%%%%%%%%%%%%%%%%%%%%%%%%%%%%%%%%%%%%%%%%%%%%%%%%%%%%%%%%%%%%%%%%%%%%%%%%%%%%%%%%%%%%%%%%%%%%
%%%%%%%%%%%%%%%%%%%%%%%%%%%%%%%%
Below we discuss how to implement semi-relativistic correction terms in  the effective  potential~(\ref{potential-CI}).% from the Lorentz structures presented in Table~\ref{table-mesones-lorentz}.  

%%%%%%%%%%%%%%%%%%%%%%%%%%%%%%%%%%%%%%%%%%%%%%%%%%%%%%%%%%%%%%%%%%%%%%%%%%%%%%%%%%%%%%%%%%%%%%%%%%%%%%%%%%%%%%%%%%%%%%%%%%%%%%%%%%%%%%%%
 %%%%%%%%%%%%%%%%%%%%%%%%%%%%%%%%%%%%%%%%%%%%%%%%%%%%%%%%%%%%%%%%%%%%%%%%%%%%%%%%%%
Dressed-quark masses and free parameters (in $\GeV$), required as input in this work are
%%%%%%%%%%%%%%%%%%%%%%%%%%%%%%%%%%%%%%%%%%%%%%%%%%%%%%%%%%%%%%%%%%%%%%%%%%%%%%%%%%
\begin{center}
\begin{tabular}{@{\extracolsep{0.5 cm}}cccc}
\hline
\hline
 $M_{\tu}=0.40$ &  $M_{\ts}=0.53$ & 
 $M_{\tc}=1.48$ & 
 $M_{\tb}=4.73$  \\
  \rule{0ex}{3.5ex}
 $\Lambda=0.905$ &
 $m_g=0.5$ &
 $\lambda=0.24$ & $\alpha=0.93\pi$\\
 \hline
 \hline
\end{tabular}
\end{center}
%%%%%%%%%%%%%%%
%%%%%%%%%%%%%%%%%%%%%%%%%%%%%%%%%%%%%%%%%%%%%%%%%%%%%%%%%%%%%%%%%%%%%%%%%%%%%%%%%%%%%%%%%%%%%%%%%%%%%%%%%%%%%%%%%%%%%%%%%%%%%%%%%%%%%%%%%%%%%%%%%%%%%%%%%%%%%%%%%%%%%%%%%%%%%%%%%%%%%%%%%%%%%
\section{Heavy-Light Mesons}
\label{HL-Mesons}
%%%%%%%%%%%%%%%%%%%%%%%%%%%%%%%%%%%%%%%%%%%%%%%%%%%%%%%%%%%%%%%%%%%%%%%%%%%%%%%%%%%%%%%%%%%%%%%%%%%%%%%%%%%%%%%%%%%%%%%%%%%%%%%%%%%%%%%%%%%%%%%%%%%%%%%%%%%%%%%%%%%%%%%%%%%%%%%%%%%%%%%%%%%%%
%\vspace{-2.5cm}
\begin{figure}[htbp]
\begin{center}
       \includegraphics[scale=0.5,angle=0]{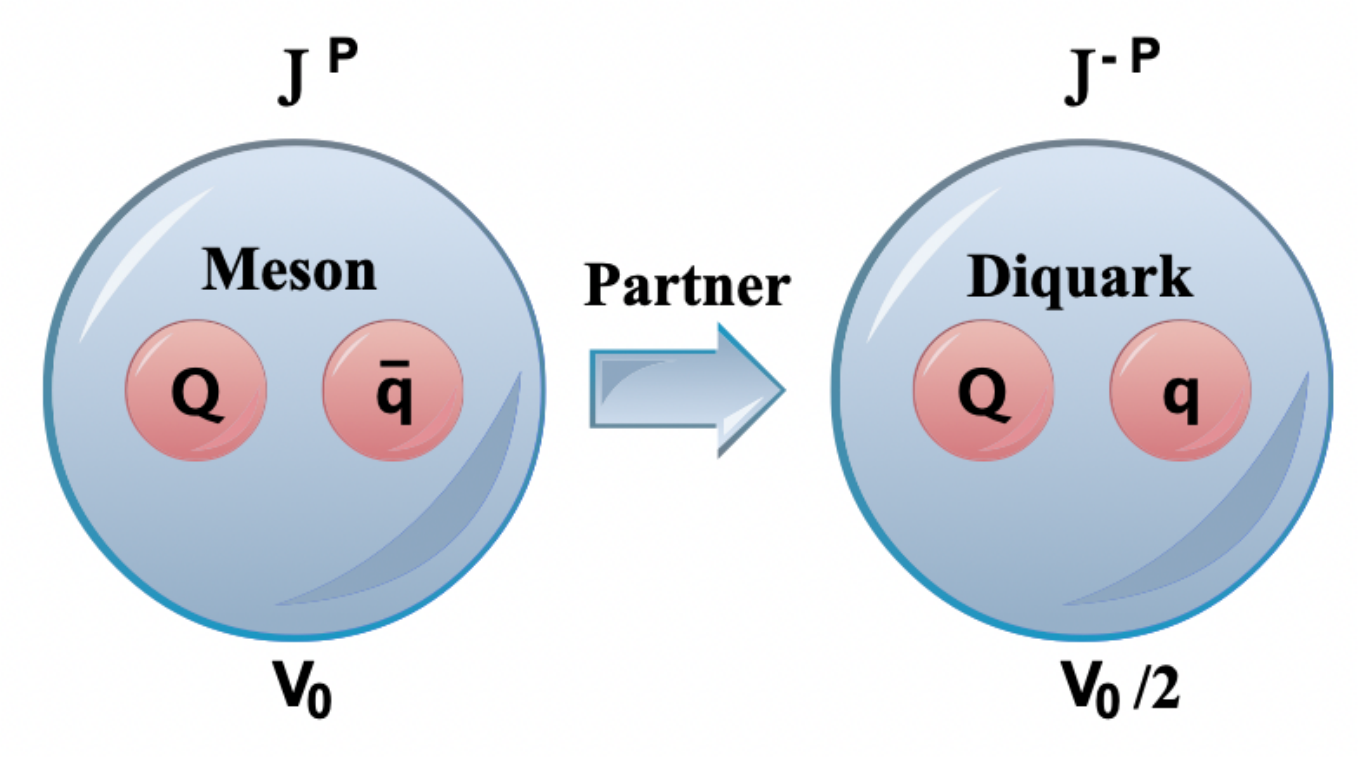}
       \caption{\label{partners} Mesons and diquarks. Here, $Q={\tc,\tb}$ and $q={\tu,\ts}$. The colour factor is diﬀerent owing to the fact that diquarks are colour antitriplets.}
       \end{center}
\end{figure}
\begin{figure}[htbp]
\begin{center}
       \includegraphics[scale=0.25,angle=-90]{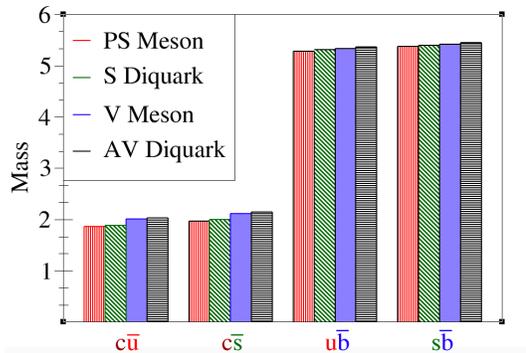}
       \caption{\label{graph3}
Mass (in GeV) of mesons and diquarks calculated with the potential $V_0$ and $\langle H_{SS0}\rangle$. It is immediate to notice that the mesons masses are smaller than for diquarks. }
       \end{center}
\end{figure}
\begin{figure}[htbp]
\begin{center}
       \includegraphics[scale=0.25,angle=-90]{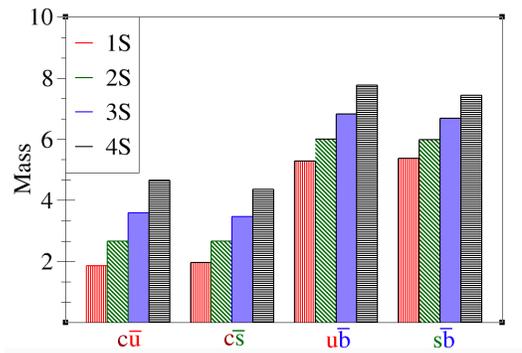}
       \caption{\label{ExPs}
Mass (in GeV) of pseudoscalar mesons and their radial excitations. }
       \end{center}
\end{figure}
\begin{figure}[htbp]
\begin{center}
       \includegraphics[scale=0.25,angle=-90]{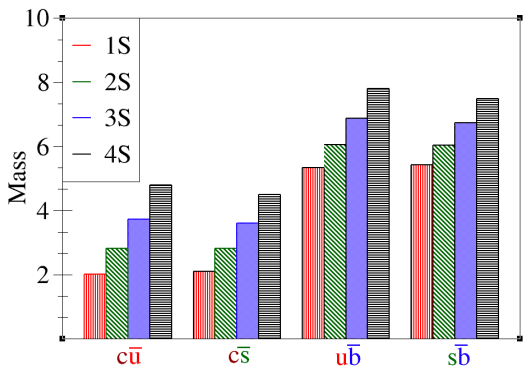}
       \caption{\label{ExVec}
Mass (in GeV) of vector mesons and their radial excitations. }
       \end{center}
\end{figure}

\%vspace{-1cm}
In order to compute the mass spectra of mesons containing one heavy quark, besides the effective potential~(\ref{potential-CI}), we also consider the spin-spin $\langle H_{SS}\rangle$ and spin-orbit $\langle H_{LS}\rangle$ interactions as~\cite{Hassanabadi:2016kqq,Lucha:1995zv}
%%%%%
\bea
V_{Qq}&=&V_0+\langle H_{SS}\rangle+\langle H_{LS}\rangle,\eea
where $\mathbf{S}$ and $\mathbf{L}$ are the total spin of the bound state and the relative orbital angular momentum of its constituents, respectively. The spin-spin term is parametrized as
\bea
H_{SS}&=&\frac{32\pi\alpha_s}{9m_qm_Q}\mathbf{S_Q\cdot S_q}|\psi(0)|^2,\eea
the relation of $|\psi(0)|^2$ for the ground state is defined as
\bea
|\psi(0)|^2=\frac{\mu}{2\pi \hbar^2}\bigg\langle \frac{dV(r)}{dr} \bigg \rangle\,.\eea
The total spin $\mathbf{S}$ is clearly given by the sum of the spins $\mathbf{S_Q}$ and $\mathbf{S_q}$. The quantum number $\mathbf{S}$  may accept precisely either of two values: $\mathbf{S}~=~0$, which corresponds to a spin singlet, {\em e.g.}  pions in the case of light quarks, or  $\eta_c$ in the charmonium system. $\mathbf{S} = 1$, corresponds to a spin triplet, {\em e.g.}  $\rho$ in the case of light quarks, or  $J/\psi$ in for charmonium \cite{Moazami:2018uqt} .
\begin{equation}
\langle H_{SS} \rangle= \left\lbrace
\begin{array}{ll}
\displaystyle{\frac{8\pi \alpha_s}{9m_q m_Q}|\psi(0)|^2}& \textup{for}\;\;\;\mathbf{S}=1,\\ \\
\displaystyle{-\frac{8\pi \alpha_s}{3m_q m_Q}|\psi(0)|^2}& \textup{for}\;\;\;\mathbf{S}=0.
\end{array}
\right.\end{equation}
The spin-orbit term has the form
\bea
H_{LS}=\left(-\frac{1}{2\mu^2 r}\frac{dV(r)}{dr}\right)\langle \mathbf{L\cdot S}\rangle.\eea
We may express the product $\mathbf{L\cdot S}$ in terms of the squares of $\mathbf{L}$, $\mathbf{S}$, and $\mathbf{J=L+S}$ as
\bea
\langle \mathbf{L\cdot S}\rangle=\frac{1}{2}[j(j+1)-\ell(\ell+1)-S(S+1)].\eea
The expectation values of the spin-orbit term vanish for $\ell=0$ or $S = 0$. Thus, in our calculations we  expect that the spin-orbit term contributes to axial mesons, but not for pseudoscalar and vector mesons.
Once we have the mass of a meson, it is immediately possible to obtain the mass of its diquark partner. The mass of a diquark with $J^P$ 
is obtained from the equation for a $J^{-P}$ meson, the difference is a factor $1/2$ of the interaction strength, see Fig.~\ref{partners}.  
Fermions and antifermions have opposite parity then, there is a change in the sign. The computed masses of mesons and diquarks with spin zero are given in Table~\ref{table-mesones-pseudo}. 
We quote the results from the effective potential $V_0$ alone  and with relativistic corrections. The values of $\langle H_{SS}\rangle$ are taken from \cite{Hassanabadi:2016kqq}. Our results are in good agreement with the last experimental measurement. $\langle H_{SS0}\rangle$ is the value that fits perfectly with the mass of the meson. $V_F$ shows the value of the diquark with the spin-spin interaction term $ \langle H_ {SS0} \rangle $.
%%%%%%%%%%%%%%%%%%%%%%%%%%%%%%%%%%%%%%%%%%%%%%%%%%%
%%%%%%%%%%%%%%%%%%%%%%%%%%%%%%%%%%%%%%%%%%%%%%%%%%%%%%%%%%%%%%%%%%%%%%%%%%%
%%%%%%%%%%%%%%%%%%%%%%%%%%%
%%%%%%%%%%%%%%%%%%%%%%%%%%%%%%%%%%%%%%%%%%%%%%%%%%%
\begin{table*}[htbp]
\caption{\label{table-mesones-pseudo}
Computed ground state masses for pseudoscalar and vector mesons and its diquarks partners (in \GeV). Experimental masses are taken from~\cite{Tanabashi:2018oca}. [.] indicates a $J^P=0^+$ diquark and \{.\} indicates a $J^P=1^+$
diquark. The column $V_0$ denotes the result from the uncorrected potential alone. $V_{OS}$ denotes the result using  $\langle H_{SS}\rangle$ from~\cite{Hassanabadi:2016kqq} and $V_F$ is the corresponding to $\langle H_{SS0}\rangle$ which provides the best fit from our model. }
%%%%%%%%%%%%%%%%%%%%%%%%%%%%%%%%%%%%%%%%%%%%%%%%%%%
%%%%%%%%%%%%%%%%%%%%%%%%%%%%%%%%%%%%%%%%%%%%%%%%%%%
\begin{center}
\begin{tabular}{@{\extracolsep{0.7 cm}}cccc|cccc|cc}
\hline
\hline
Pseudoscalar &&&& Scalar\\
\hline
Mesons   & Exp. & $V_0$& $V_{0S}$ & Diquark & $V_0$&$V_{0S}$&$V_{F}$&$\langle H_{SS}\rangle $& $\langle H_{SS0}\rangle $\\
%%%%%%%%%%%%%%%%%%%%%%%%%%%%%%%%%%%%%%%%%%
$D^{0}(\tc\bar{\tu})$  &  1.86& 2.06& 1.98 &$[\tc\tu]$& 2.07&2.00&1.88&-0.080&-0.19 \\%2.06  ,  1.98  ,  2.07  ,  1.99  ,  1.88
%%%%%%%%%%%%%%%%%%%%%%%%%%%%%%%%%%%%%%%%%%%%%%%%
$D^{+}_{\ts}(\tc\bar{\ts})$  & 1.97 &2.14& 2.08&$[\tc\ts]$& 2.17&2.11&2.0&-0.056&-0.17\\%2.19  ,  2.08  ,  2.20  ,  2.11  ,  2
%%%%%%%%%%%%%%%%%%%%%%%%%%%%%%%%%%%%%%%%%%%%%%%%%%%
$B^{+} (\tu\bar{\tb})$ & 5.28 &5.27&5.25& $[\tu\tb]$&5.30&5.27&5.31&-0.025&0.01\\ %5.27  ,  5.25  ,  5.30  ,  5.27  ,  5.31
%%%%%%%%%%%%%%%%%%%%%%%%%%%%%%%%%%%%%%%%%%%%%%%%%%%%
$B_s^0(\ts\bar{\tb})$ & 5.37 &5.35&5.33&$[\ts\tb]$& 5.38&5.37&5.40&-0.017&0.02\\%5.35  ,  5.33  ,  5.43  ,  5.37  ,  5.4
\hline
\hline
Vector &&&& Vector-Axial\\
\hline
Mesons   & Exp. &$V_C$& $V_{0S}$ & Diquark && $V_{0S}$&$V_F$&$\langle H_{SS}\rangle$ &$ \langle H_{SS0}\rangle $\\
$D^{*0}(\tc\bar{\tu})$& 2.01&2.02&2.08&$\{\tc\tu\}$&&2.10&2.03&0.026& -0.045  \\%2.08  ,  2.1
$D_{\ts}^{*}(\tc\ts)$ & 2.11&2.57&2.16&$\{\tc\ts\}$&&2.18&2.14&0.018&-0.025\\%2.16  ,  2.18
$B^{+*}(\tu\bar{\tb})$ &5.33&5.66&5.28&$\{\tu\tb\}$&&5.30&5.36&0.008&0.06 \\%5.28  ,  5.3
$B_{\ts}^{0*}(\ts\bar{\tb})$ & 5.42&5.38&5.36&$\{\ts\tb\}$&&5.39&5.45&0.005&0.07\\%5.36  ,  5.39
\hline
\hline
\end{tabular}
\end{center}
\end{table*}
%%%%%%%%%%%%%%%%%%%%%%%%%%%%%%%%%%%%%%%%%%%%%%%%%%%%%%%%%%%%%%%%%%%%%%%%%%%%%%%%%%
\begin{table*}[ht]
\caption{\label{diquarks-comp}Diquark differences with other models}
\begin{center}
\begin{tabular}{@{\extracolsep{0.5 cm}}ccccccccc}
\hline
\hline
&$[\tc\tu]$&$[\tc\ts]$&$[\tu\tb]$&$[\ts\tb]$&\{\tc\tu\}&\{\tc\ts\}&\{\tu\tb\}&\{\ts\tb\}\\
Our& 1.88&2.0&5.31&5.40&2.03&2.14&5.36&5.45\\
%%%%%%%%%%%%%%%%%%%%%%%%%%%%%%%%%%%%%%%%%%%%
Ref. \cite{Gutierrez-Guerrero:2019uwa}&2.01&2.13&5.23&5.34&2.09&2.19&5.26&5.36\\
%%%%%%%%%%%%%%%%%%%%%%%%%%%%%%%%%%%%%%%%%%%%%
Diff.& 6.46\%&6.10\%&1.53\%&1.12\%&2.87\%&2.28\%&1.90\%&1.67\%\\
%%%%%%%%%%%%%%%%%%%%%%%%%%%%%%%%%%%%%%%%%%%%%%%
Ref.\cite{Yin:2019bxe}&2.15&2.26&5.51&5.60&2.24&2.34&5.53&5.62\\
%%%%%%%%%%%%%%%%%%%%%%%%%%%%%%%%%%%%%%%%%%%%%%%%
Diff.&12.55\%&11.50\%&3.63\%&3.57\%&9.37\%&8.55\%&3.07\%&3.02\%\\
\hline
\hline
\end{tabular}
\end{center}
\end{table*}
%%%%%%%%%%%%%%%%%%%%%%%%%%%%%%%%%%%%%%%%%%%%%%%%%%%%%%%%%%%%%%%%%%%%%%%%%%%%%%%%%%%%%%%%%%%%%%%%%%%%%%%%%%%%%%%%%%%%%%%%%%%%%%%%%%%%%%%%%%%%%%%%%%%%%%%%
%%%%%%%%%%%%%%%%%%%%%%%%%%%%%%%%%%%%%%%%%%%%%%%%%%%%%%%%%%%%%%%%%%%%%%%%%%%%%%%%%%%%%%%%%%%%%%%%%%%%%%%%%%%%%%%%%%%%%%%%%%%%%%%%%%%%%%%%%%%%%%%%%%%%%%%%%%%%%%%%%%%%%%%%%%%%%%%%%%%%%%%%%%%%%
%It is plain from Fig.~\ref{graph3} that the level ordering of %diquark correlations is precisely the same as that for mesons. % The mass of the diquark's partner meson is a fair guide to the %diquark’s mass.
From Fig.~\ref{graph3} it is clear that the level ordering of diquark masses is exactly the same as that of the mesons. The mass of the mesons is a reliable guide to predict the mass of the diquarks.
%%%%%%%%%%%%%%%%%%%%%%%%%%%%%%%%%%
%%%%%%%%%%%%%%%%%%%%%%%%%%%%%%%%%%%%%%%%%%%%%%%%%%%%
Table~\ref{diquarks-comp} shows a comparison between our results  and other findings.  The calculation of masses of ground-state baryons with positive parity in the quark-diquark picture requires scalar $0^+$ and axial-vector $1^+$ diquarks. This calculation will serve in future works to calculate the masses of baryons using this model. In Table~\ref{table-HL-Ex} we show our prediction for radially excited Heavy-Light mesons mass. Charm and charm-strange mesons are compared with the results in~\cite{Godfrey:2015dva} and bottom  and  bottom-strange mesons are compared with~\cite{Lu:2016bbk}.\\
%%%%%%%%%%%%%%%%%%%%%%%%%%%%%%%%%%%%%%%%%%%%%%%%%%%
%%%%%%%%%%%%%%%%%%%%%%%%%%%%%%%%%%%%%%%%%%%%%%%%%%%
%The wave functions of the ground state for pseudoscalar and vector mesons are plotted in Figs.~\ref{mesones-HL-PS.pdf} and~\ref{mesonesVS1.pdf}, respectively. We can see the effect of the $\langle H_{SS0}\rangle$ factor, it makes the sign of the wave function change since it is the only difference between the pseudoscalar mesons and the vector ones.
 Considering the known mass difference of the ground state pseudoscalar and vector mesons,
\bea
\nn m_{D^{0}}-m_{D^{*0}}&=&150\;\textup{MeV}\;,\\
\nn m_{D^{+}_{\ts}}-m_{D^{*}_{\ts}}&=&140\;\textup{MeV}\;,\\
\nn m_{B^{+}}-m_{B^{+*}}&=&50\;\textup{MeV}\;,\\
\nn m_{B^{0}_{\ts}}-m_{B^{0*}_{\ts}}&=& 50\;\textup{MeV}\;,\eea
%%%%%%%%%%%%%%%%%%%%%%%%%%%
we observe that the mass-splitting of the  resonances predicted by our model in the Heavy-Light sector, Table \ref{table-HL-Ex}, are generally $140 \sim 150\;\textup{MeV}$ for the case of mesons containing a quark $\tc$, whereas for $b$-mesons, the difference is approximately $40$ to $50$ MeV. In the $b$-flavored meson sector, experimental data for excited $b$-meson states are limited for now. But still several $b$-flavored mesons have been observed~\cite{Chen:2016spr}. In Figs.~\ref{ExPs} and~\ref{ExVec} we emphasize the difference between the excited states and the ground state, the percentage differences are given in Table~\ref{difEx}.\\
\begin{table}[ht]
\caption{\label{difEx} Percentage differences between ground state and radial excitations for Heavy-Light mesons. }
%%%%%%%%%%%%%%%%%%%%%%%%%%%%%%%%%%%%%%%%%%%%%%%%%%%
%%%%%%%%%%%%%%%%%%%%%%%%%%%%%%%%%%%%%%%%%%%%%%%%%%%
\begin{center}
\begin{tabular}{@{\extracolsep{0.3 cm}}cccc}
\hline
\hline
PS Meson &2S&3S&4S\\
$D^{0}(\tc\bar{\tu})$&43.55\%&93.01\%&150.54\%\\
%%%%%%%%%%%%%%%%%%%%%%%%%%%%%%%%%%%%%%%%%%%%%%%%
$D^{+}_{\ts}(\tc\bar{\ts})$&35.53\%&76.14\%& 120.81\% \\
%%%%%%%%%%%%%%%%%%%%%%%%%%%%%%%%%%%%%%%%%%%%%%%%%%%
$B^{+} (\tu\bar{\tb})$&13.82\%&29.36\%&46.97\% \\
%%%%%%%%%%%%%%%%%%%%%%%%%%%%%%%%%%%%%%%%%%%%%%%%%%%%
$B_s^0(\ts\bar{\tb})$&11.54\%&24.58\%&38.55\% \\
\hline
\hline
V Meson &2S&3S&4S\\
$D^{*0}(\tc\bar{\tu})$&40.30\% &85.57\%&138.80\% \\
$D_{\ts}^{*}(\tc\bar{\ts})$&33.65\%&71.09\%&113.27\% \\
$B^{+*}(\tu\bar{\tb})$&13.70\%&29.08\%&46.53\% \\
$B_{\ts}^{0*}(\ts\bar{\tb})$&11.25\%&24.35\%&38.19\% \\
\hline
\hline
\end{tabular}
\end{center}
\end{table}
In the diquark sector, the differences between scalar and axial vector diquarks are larger than for mesons. The masses predicted in this work for diquarks in their ground-state are smaller than those presented in~\cite{Yin:2019bxe} and they are very close to those obtained in~\cite{Gutierrez-Guerrero:2019uwa} using a field theoretical approach based on the Schwinger-Dyson equations.
%%%%%%%%%%%%%%%%%%%%%%%%
%\begin{figure}[ht]
%       \includegraphics[scale=0.25,angle=-90]{mesones-HL-PS.pdf}
%       \caption{\label{mesones-HL-PS.pdf}
%The $1S$ state for Heavy-Light Pseudoescalar Mesons. $D^0$, $B^+$ have wave functions positive and $D_{\ts}^+$, $B_{\ts}^{0}$ are negative.}
%\end{figure}
%%%%%%%%%%%%%%%%%%%%%%%%%%%%%%%%%%%%%%%%%%%%%%%%%%%%%%%%%%%%%%%%%%%%%%%%%%%%%%%%%%%%%%
%\begin{figure}[ht]
%\begin{center}
%       \includegraphics[scale=0.25,angle=-90]{MesonesVS1.pdf}
%       \caption{\label{mesonesVS1.pdf}
%The $1S$ state for Heavy-Light Vector Mesons. $D^{0*}$, $B^{+*}$ have wave functions negative and $D_{\ts}^*$, $B_{\ts}^{0*}$ are positive.}
%       \end{center}
%\end{figure}
%%%%%%%%%%%%%%%%%%%%%%%%%%%%%%%%%%%%%%%%%%%%%%%%%%%%%%%%%%%%%%%%%%%%%%%%%%%%%%%%%%
%%%%%%%%%%%%%%%%%%%%%%%%%%%%%%%%%%%%%%%%%%%%%%%%%%%
%%%%%%%%%%%%%%%%%%%%%%%%%%%%%%%%%%%%%%%%%%%%%%%%%%%
\begin{table*}[htp]
\caption{\label{table-HL-Ex}
Computed masses for excited pseudoscalar an vector mesons and its diquark partners (in \GeV). For convenience, we introduce the shorthand notation $n^{2S+1}L_J$, where $n$ stands for the principal quantum number.  We take $N=100$.}
%%%%%%%%%%%%%%%%%%%%%%%%%%%%%%%%%%%%%%%%%%%%%%%%%%%
%%%%%%%%%%%%%%%%%%%%%%%%%%%%%%%%%%%%%%%%%%%%%%%%%%%
\begin{center}
\begin{tabular}{@{\extracolsep{0.7 cm}}cccc|cccc}
\hline
\hline
Mesons & Ref. \cite{Godfrey:2015dva} & $V_F$& Diquark&Mesons & Ref. \cite{Godfrey:2015dva} & $V_F$& Diquark\\ 
$2\;^1S_0(\tc\bar{\tu})$&2.58&2.67&2.53& $2\;^3S_1(\tc\bar{\tu})$&2.64&2.82&2.67
\\
$3\;^1S_0(\tc\bar{\tu})$&3.06&3.59&3.33
&$3\;^3S_1(\tc\bar{\tu})$&3.11&3.73&3.47
\\
$4\;^1S_0(\tc\bar{\tu})$&3.47&4.66&4.39
&$4\;^3S_1(\tc\bar{\tu})$&3.50&4.80&4.53
\\
%%%%%%%%%%%%%%%%%%%%%%%%%%%%%%%%%%%%%%%%%%%%%%%%%%%
$2\;^1S_0(\tc\bar{\ts})$&2.67&2.67&2.55
&$2\;^3S_1(\tc\bar{\ts})$&2.73&2.82&2.69
\\
$3\;^1S_0(\tc\bar{\ts})$&3.15&3.47&3.21
&$3\;^3S_1(\tc\bar{\ts})$&3.19&3.61&3.36
\\
$4\;^1S_0(\tc\bar{\ts})$&3.55&4.35&4.07
&$4\;^3S_1(\tc\bar{\ts})$&3.58&4.50&4.22
\\
\hline
& Ref. \cite{Lu:2016bbk}&&&&Ref. \cite{Lu:2016bbk}&&\\
%%%%%%%%%%%%%%%%%%%%%%%%%%%%%%%%%%%%%%%%%%%%%%%%%
$2\;^1S_0(\tu\bar{\tb})$&5.91&6.01&5.88
& $2\;^3S_1(\tu\bar{\tb})$&5.94&6.06&5.93
\\
$3\;^1S_0(\tu\bar{\tb})$&6.37&6.83&6.57
& $3\;^3S_1(\tu\bar{\tb})$&6.40&6.88&6.62
\\
$4\;^1S_0(\tu\bar{\tb})$&--&7.76&7.48
& $4\;^3S_1(\tu\bar{\tb})$&--&7.81&7.53
\\
%%%%%%%%%%%%%%%%%%%%%%%%%%%%%%%%%%%%%%%%%%%%%%%%%%%
$2\;^1S_0(\ts\bar{\tb})$&5.98&5.99&5.89
& $2\;^3S_1(\ts\bar{\tb})$&6.00&6.03&5.94
\\
$3\;^1S_0(\ts\bar{\tb})$&6.42&6.69&6.45
& $3\;^3S_1(\ts\bar{\tb})$&6.44&6.74&6.50
\\
$4\;^1S_0(\ts\bar{\tb})$&--&7.44&7.16
& $4\;^3S_1(\ts\bar{\tb})$&--&7.49&7.21
\\
%%%%%%%%%%%%%%%%%%%%%%%%%%%%%%%%%%%%%%%%
%%%%%%%%%%%%%%%%%%%%%%%%%%%%%%%%%%%%%%%%%%%%%%%%%%%
%%%%%%%%%%%%%%%%%%%%%%%%%%%%%%%%%%%%%%%%%%%%%%%%%%%
\hline
\hline
\end{tabular}
\end{center}
\end{table*}
%%%%%%%%%%%%%%%%%%%%%%%%%%%%%%%%%%%%%%%%
%%%%%%%%%%%%%%%%%%%%%%%%%%%%%%%%%%%%%%%%%%%%%%%%%%%%%%%%%%%%%%%%%%%%%%%%%%%%%%%%%%%%%%%%%%%%%%%%%%%%%%%%%%%%%%%%%%%%%%%%%%%%%%%%%%%%%%%%%%%%%%%%%%%%%%%%%%%%%%%%%%%%%%%%%%%%%%%%%%%%%%%%%%
%%%%%%%%%%%%%%%%%%%%%%%%%%%%%%%%%%%%%%%%%%%%%%%%%%%%
%%%%%%%%%%%%%%%%%%%%%%%%%%%%%%%%%%%%%%%%%%%%%%%%%%%%%%%%%%%%%%%%%%%%%%%%%%%%%%%%%%%%%%%%%%%%%%%%%%%%%%%%%%%%%%%%%%%%%%%%%%%%%%%%%%%%%%%%%%%%%%%%%%%%%%%%%%%%%%%%%%%%%%%%%%%%%%%%%%%%%%%%%%%%%
%%%%%%%%%%%%%%%%%%%%%%%%%%%%%%%%%%%%%%%%%%%%%%%%%%%
%%%%%%%%%%%%%%%%%%%%%%%%%%%%%%%%%%%%%%%%%%%%%%%%%%%
\section{Heavy Mesons}
\label{Heavy-Mesons}
\begin{figure}[H]
\begin{center}
       \includegraphics[scale=0.25,angle=-90]{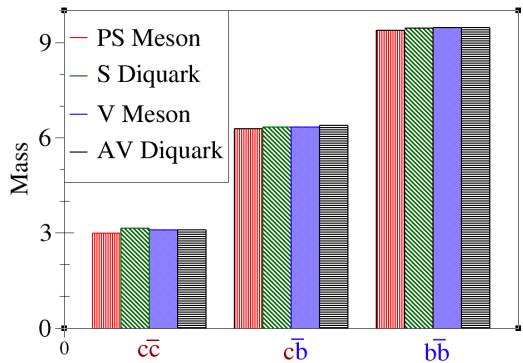}
       \caption{\label{dimesonH}
Mass of mesons and diquarks calculated with the potential $V_F$ and $\langle H_{SS0}\rangle$. The largest difference between a meson and its diquark partner occurs in the $\eta_{\tc}$ meson, and is approximately 5\%.The mass splitting between the other heavy mesons and their diquarks does not exceed 1\%. }
       \end{center}
\end{figure}

%%%%%%%%%%%%%%%%%%%%%%%%%%%%%%%%%%%%%%%%
%%%%%%%%%%%%%%%%%%%%%%%%%%%%%%%%%%%%%%%%%%%%%%%%%%%
%%%%%%%%%%%%%%%%%%%%%%%%%%%%%%%%%%%%%%%%%%%%%%%%%%%
In what follows, we use the same line of thoughts used in the previous Section to calculate the heaviest mesons $\tb\tb$, $\tc\tc$ and $\tc\tb$. Our results for ground and excited states of charmonia and bottomonia are reported in Table~\ref{CYB}.
%%%%%%%%%%%%%%%%%%%%%%%%%%%%%%%%%%%%%%%%%%%%%%%%%%%%%%%%%%%%%%%%%%%%%%%%%%%%%%%%%%%%%%%%%%%%%%%%%%%%%%%%%%%%%%%%%%%%%%%%%%%%%%%%%%%%%%%
\begin{center}
\begin{table*}[ht]
\centering
\caption{\label{CYB}Heavy mesons masses (in GeV) for potential $V_0$ compared
to the experimental masses from PDG. By taking N=100 $r_{max}=10$fm.} %%%%%%%%%%%%%%%%%%%%%%%%%%%%%%%%%%%%%%%%%%%%%%%%%%%
\begin{tabular}[t]{@{\extracolsep{0.7 cm}}cccc|cccc}
\hline
\hline
$\;$ State $\;$ & Exp. \cite{Agashe:2014kda} & $V_0$ & Diquark&State& Exp. \cite{Agashe:2014kda} &$V_F$& Diquark\\
$1\hspace{0.1cm}^1S_0(\tc\bar{\tc})$ & $2.98$ & $2.98$ & 3.14&$1\hspace{0.1cm}^3S_1 (\tc\bar{\tc})$ &3.10&3.10&3.15\\
%%%%%%%%%%%%%%%%%%%%%%%%%%%%%%%%%
$2\hspace{0.1cm}^1S_0 (\tc\bar{\tc})$ & $3.64$ & $3.45$ & 3.65 &$2\hspace{0.1cm}^3S_1 (\tc\bar{\tc})$&3.69&3.57&3.51 \\
%%%%%%%%%%%%%%%%%%%%%%%%%%%%%%%%
$3\hspace{0.1cm}^1S_0(\tc\bar{\tc})$ & $-$ & $3.98$  & 4.20& $3\hspace{0.1cm}^3S_1 (\tc\bar{\tc})$&4.04&4.10&3.92 \\
%%%%%%%%%%%%%%%%%%%%%%%%%%%%%%%%%%
$4\hspace{0.1cm}^1S_0(\tc\bar{\tc})$ & $-$ & $4.55$ & 4.89&
$4\hspace{0.1cm}^3S_1 (\tc\bar{\tc})$&4.42&4.67&4.39
%%%%%%%%%%%%%%%%%%%%%%%%%%%%%%%%%%
\\
$1\hspace{0.1cm}^1S_0 (\tb\bar{\tb})$ & $9.40$ & $9.38$ & 9.45
&$1\hspace{0.1cm}^3S_1 (\tb\bar{\tb})$&9.46&9.46&9.53\\
%%%%%%%%%%%%%%%%%%%%%%%%%%%%%%%%%%
$2\hspace{0.1cm}^1S_0 (\tb\bar{\tb})$ & $9.99$ & $9.60$ & 9.62
&$2\hspace{0.1cm}^3S_1 (\tb\bar{\tb})$&10.02&9.68&9.70\\
%%%%%%%%%%%%%%%%%%%%%%%%%%%%%%%%%%
$3\hspace{0.1cm}^1S_0 (\tb\bar{\tb})$ & $-$ & $9.85$  & 9.81
&$3\hspace{0.1cm}^3S_1 (\tb\bar{\tb})$&10.36&9.93&9.89 \\
%%%%%%%%%%%%%%%%%%%%%%%%%%%%%%%%%%
$4\hspace{0.1cm}^1S_0 (\tb\bar{\tb}) $& $-$ & $10.13$ & 10.03  
&$4\hspace{0.1cm}^3S_1 (\tb\bar{\tb})$&10.58&10.21&10.11
\\
\hline
%%%%%%%%%%%%%%%%%%%%%%%%%%%%%%%%%%%%%%%%%%%%%%%%%%%
State& Ref. \cite{Li:2019tbn}&$V_F$&Diquark&State&Ref. \cite{Li:2019tbn}&$V_F$&Diquark\\
$1\;^1S_0(\tc\bar{\tb})$&6.27&6.27&6.33&
$2\;^3S_1(\tc\bar{\tb})$&6.33&6.33&6.38\\
%%%%%%%%%%%%%%%%%%%%%%%%%%%%%%%%%%%
$2\;^1S_0(\tc\bar{\tb})$&6.87&6.62&6.60& $2\;^3S_1(\tc\bar{\tb})$&6.89&6.68&6.66
\\
$3\;^1S_0(\tc\bar{\tb})$&7.24&7.03&6.92& $3\;^3S_1(\tc\bar{\tb})$&7.25&7.09&6.97
\\
$4\;^1S_0(\tc\bar{\tb})$&7.54&7.48&7.25& $4\;^3S_1(\tc\bar{\tb})$&7.55&7.54&7.30
\\
%%%%%%%%%%%%%%%%%%%%%%%%%%%%%%%%%%%%%%%%%%%%%%%%%%%
\hline
\hline
\end{tabular}
\end{table*}
\end{center}
%%%%%%%%%%%%%%%%%%%%%%%%%%%%%%%%%%%%%%%%%%%%%%%%%%%%%%%%%%%%%%%%%%%%%%%%%%%%%%%%%%
%%%%%%%%%%%%%%%%%%%%%%%%%%%%%%%%%%%%%%%%%%%%%%%%%%%%%%%%%%%%%%%%%%%%%%%%%%%%%%%%%%%%%%%%%%%%%%%%%%%%%%%%%%%%%
We use the following values for $\langle H_{SS0}\rangle$,
\begin{center}
\begin{tabular}{@{\extracolsep{0.5 cm}}ccc}
\hline
\hline
 Meson & Pseudoscalar& Vector \\
 $\tc\bar{\tc}$ & 0 &0.12\\
 $\tb\bar{\tb}$ & 0 &0.085\\
 $\tc\bar{\tb}$ & 0.085 & 0.14\\
 \hline
 \hline
\end{tabular}
\end{center}
As in the case of Heavy-Light mesons, we calculate the mass of diquark partners that contain two heavy quarks. In Fig.~\ref{dimesonH} we show the mass splitting of mesons and diquarks.
\\
We compare these results with existing ones using other models in Table~\ref{di-H}.
\begin{table}
\caption{\label{di-H} Results for diquark masses (in GeV). Second and fifth rows contain the difference between the masses obtained herein and those obtained in \cite{Gutierrez-Guerrero:2019uwa},\cite{Yin:2019bxe}.}
\begin{center}
\begin{tabular}{@{\extracolsep{0.2 cm}}ccccccc}
\hline
\hline
  & $[\tc\tc]$ & $\{\tc\tc\}$ &  $ [\tb\tb]$&$ \{\tb\tb\}$&$ [\tc\tb]$& $\{\tc\tb\}$\\
  Herein&3.14&3.15&9.45&9.53&6.33&6.38\\
  Ref. \cite{Gutierrez-Guerrero:2019uwa}&3.11&3.12&9.53&9.53&6.31&6.31\\
 Diff.&0.96\%&0.96\%&0.83\%&0\%&0.31\%&1.10\%\\
 Ref. \cite{Yin:2019bxe}&--&3.30&--&9.68&6.48&6.50\\
  Diff.&--&4.54\%&--&1.54\%&2.31\%&1.84\%\\
 \hline
 \hline
\end{tabular}
\end{center}
\end{table}
%%%%%%%%%%%%%%%%%%%%%%%%%%%%%%%%%%%%%%%%%%%%%%%%%%%%%%%%%%%%%%%%%%%%%%%%%%%%%%%%%%%%%%%%%%%%%%%%%%%%%%%%%%%%%%%%%%%%%%%%%%%%%%%%%%%%%%%%%%%%%%%%%%%%%%%%%%%%%%%%%%%%%
The hyperfine mass-splitting of singlet-triplet states
\bea\Delta m_{hf}= m(n\;^{3}S_1)- m(n\;^{1}S_0)\,,\eea
probes the spin dependence of bound-state energy levels and imposes constraints on theoretical descriptions \cite{Segovia:2016xqb}. The hyperfine mass splitting  predicted by our  model are $120$ MeV, $80$ MeV and $60$ MeV for the  different flavours mesons $\tc\bar{\tc}$, $\tb\bar{\tb}$ and  $\tc\bar{\tb}$.
%%%%%%%%%%%%%%%%%%%%%%%%%%%%%%%%%%%%%%%%%%%%%%%%%%%%%%%%%%%%%%%%%%%%%%%%%%%%%%%%%%
%%%%%%%%%%%%%%%%%%%%%%%%%%%%%%%%%%%%%%%%%%%%%%%%%%%%%%%%%%%%%%%%%%%%%%%%%%%%%%%%%%%%%%%%%%%%%%%%%%%%%%%%%%%%%%%%%%%%%%%%%%%%%%%%%%%%%%%
%%%%%%%%%%%%%%%%%%%%%%%%%%%%%%%%%%%%%%%%%%%%%%%%%%%%%%%%%%%%%%%%%%%%%%%%%%%%%%%%%%
%%%%%%%%%%%%%%%%%%%%%%%%%%%%%%%%%%%%%%%%%%%%%%%%%%%%%%%%%%%%%%%%%%%%%%%%%%%%%%%%%%%%%%%%%%%%%%%%%%%%%%%%%%%%%%%%%%%%%%%%%%%%%%%%%%%%%%%%%%%%%%%%%%%%%%%%%
Recently, CMS Collaboration~\cite{Sirunyan:2019osb} observed two peaks for the excited states of $B_\tc$ meson, $B^{+}_{\tc}(2\;^1S_0)$ and $B^{+*}_{\tc}(2\;^3S_1)$. The mass of $B^{+}_{\tc}(2S)$ is measured to be $6.871$ GeV \cite{Sirunyan:2019osb}. A mass difference of $29.1 \pm 1.5$ MeV is measured between two states. However exact mass of $B^{+*}_{\tc}(2S)$ is unknown. The mass  difference that we find with this potential is $60$ MeV for all radial excitations of $B_\tc$.
%%%%%%%%%%%%%%%%%%%%%%%%%%%%%%%%%%%%%%%%%%%%%%%%%%%%%%%%%%%%%%%%%%%%%%%%%%%%%%%%%%%%%%%%%%%%%%%%%%%%%%%%%%%%%
%%%%%%%%%%%%%%%%%%%%%%%%%%%%%%%%%%%%%%%%%%%%%%%%%%%%
%%%%%%%%%%%%%%%%%%%%%%%%%%%%%%%%%%%%%%%%%%%%%%%%%%%%%%%%%%%%%%%%%%%%%%%%%%%%

%\begin{figure}[ht]
%\begin{center}
%       \includegraphics[scale=0.25,angle=-90]{CB-PS.pdf}
%       \caption{Wavefunctions of $S$-states for $B_{\tc}^{+}$ mesons.}
%       \end{center}
%\end{figure}
%%%%%%%%%%%%%%%%%%%%%%%%%%%%%%%%%%%%%%%%%%%%%%%%%%%%%%%
%\begin{figure}[ht]
%\begin{center}
%       \includegraphics[scale=0.25,angle=-90]{CB-V.pdf}
%       \caption{Wavefunctions of $S$-states for $B_\tc^{*}$ mesons.}
%       \end{center}
%\end{figure}

\section{Conclusions}
\label{conclusions}
%%%%%%%%%%%%%%%%%%%%%%%%%%%%%%%%%%%%%%%%%%%%%%%%%%%%%%%%%%%%%%%%%%%%%%%%%%%%%%%%%%%%%%%%%%%%%%%%%%%%%%%%%%%%%%%%%%%%%%%%%%%%%%%%%%%%%%%%%%%%%%%%%%%%%%%%%%%%%%%%%%%%%
In this article, we present results of the calculations of the mass spectra  Heavy-Heavy ($\eta_\tc$, $\eta_\tb$, $B_\tc$,  $J/\psi$, $\Upsilon$) and Heavy-Light mesons ($D$, $D_\ts$, $B$, $B_\ts$) in Tables~\ref{table-mesones-pseudo} and~\ref{CYB}, respectively,  and their radial excitations in Tables~\ref{table-HL-Ex} and~\ref{CYB} along with their associated diquarks within a non-relativistic framework. We recall an important property of Heavy-Light mesons, which is that in the infinite heavy quark mass limit, the properties of the meson are determined by those of the light quark. For our task, we solved a one-body reduced Schrödinger equation through the MNM, hence converting the problem of determining the masses of these mesons as the eigenvalues of a matrix. Our key modification was the introduction of the interaction potential. We proposed an effective form of the non-relativistic potential from a symmetry preserving Poincaré covariant vector-vector CI model of QCD~\cite{GutierrezGuerrero:2010md}. In order to present the most transparent analysis possible, we followed Refs.~\cite{Hassanabadi:2016kqq,alfredo,alfredo2021} and we included spin-orbit and spin-spin interaction terms and  predict the mass spectra  of Heavy-Heavy and Heavy-Light mesons, along with their corresponding diquarks. At this level, the equation for a $J^P$ diquark is readily obtained from that for a $J^{-P}$ meson. Our findings are in agreement with experimental results and other theoretical calculations.  In particular, the mass spectra of diquarks is closely related to the mass of the meson itself. We would like to emphasise here that the mass differences of meson states, which are smaller than the experimental measurements is consistent with the findings of Ref.~\cite{Serna:2017nlr}, where the interaction of Heavy and Light quarks is taken in the same footing. A modification of the interaction strength when light quarks are involved (see, for instance, Ref.~\cite{Serna:2020txe}) is currently under consideration. Nevertheless, we emphasize that our  study lays the foundation in this model to explain recently discovered Heavy-Light tetraquarks states, which can be considered to have an internal structure consisting in diquarks and antidiquarks. Furthermore, an estimate of the mass of baryonic systems as a quark-diquark bound state is currently under consideration.

%as well as other multiquark systems is currently under consideration. Findings shall be report elsewhere.

\begin{acknowledgements}
LXGG acknowledges financial support by CONACyT under the program “Cátedras-CONACyT".  We acknowledge Bruno El-Bennich and Khépani Raya for valuable discussions and careful reading of the manuscript.
\end{acknowledgements}
\bibliography{ccc-a}
\end{document}